\documentclass[12pt.preprint]{aastex}

\usepackage{graphicx}
\shorttitle{Smith's Cloud}

\begin{document}
\title{Smith's Cloud: a High-velocity Cloud Colliding with the Milky Way}

\author{Felix J. Lockman}
\affil{National Radio Astronomy Observatory\altaffilmark{1},
  P.O. Box 2, Green Bank, WV, 24944, jlockman@nrao.edu}

\author{Robert A. Benjamin \& A.J. Heroux}
\affil{University of Wisconsin - Whitewater, 
  800 W. Main St, Whitewater, WI 53190, benjamir@uww.edu, 
ajh8v@cms.mail.virginia.edu}

\author{Glen I. Langston}
\affil{National Radio Astronomy Observatory,
  P.O. Box 2, Green Bank, WV, 24944, glangston@nrao.edu}

\email{jlockman@nrao.edu}

\altaffiltext{1}{The National Radio Astronomy Observatory is operated
by Associated Universities, Inc., under cooperative agreement with the
National Science Foundation.}

\begin{abstract} 

New 21cm H\,I observations made with the Green Bank Telescope show
that the high-velocity cloud known as Smith's Cloud has a striking
cometary appearance and many indications of interaction with the
Galactic ISM.  The velocities of interaction give a kinematic distance
of $12.4\pm1.3$ kpc, consistent with the distance derived from other
methods.  The Cloud is $>3 \times 1$ kpc in size and its tip at 
$(\ell,b) \approx 39\arcdeg-13\arcdeg$ is 7.6 kpc from the Galactic center
and 2.9 kpc below the Galactic plane. It has 
$>10^6$ M$_{\Sun}$ in  H\,I.   Its
leading section has a total space velocity near 300 km s$^{-1}$, 
is moving toward the Galactic plane with a velocity of $73\pm26$ km
s$^{-1}$,  and  is shedding material to the Galaxy.  
In the absence of drag the Cloud will cross the plane in
about 27 Myr.  Smith's Cloud may be an example of the accretion of
gas by the Milky Way needed to explain certain persistant
anomalies in Galactic chemical evolution.

\end{abstract}

\keywords{Galaxy: halo -- Galaxy: evolution -- ISM: Clouds  
--- ISM: individual (Smith's Cloud)}

\section{Introduction}

High-velocity Clouds (HVCs)  cover as much as 40\% of the sky and have
been hypothesized to be the remnants of the formation of the Milky
Way, products of a Galactic fountain, material stripped from the
Magellanic Clouds, satellites of the Milky Way, and objects in the Local Group 
 \citep{WvW1997, Blitzetal1999, BraunBurton1999, 
Lockmanetal2002, MallerBullock2004, Connorsetal2006}.  Several 
 have distance determinations which place them in the halo of
the Galaxy with  $M_{\rm HI} \sim 10^6  - 10^7 M_{\Sun}$ 
\citep{Thometal2007, Wakkeretal2007, Wakkeretal2008}.  Some 
 have a cometary morphology and kinematics 
suggesting that they are interacting with an external medium
\citep{MirabelMorras1990,Bruns2000, BrunsMebold2004, Peeketal2007}.

Smith's Cloud \citep{Smith1963} is a large, coherent H\,I feature 
also called the Galactic Center Positive (GCP) complex
\citep{WvW1997}. Its velocity of $ +100$ km~s$^{-1}$ is
only slightly larger than permitted by Galactic rotation at its
location (${\ell,b \approx 39\arcdeg,-13\arcdeg}$), and Smith
concluded that it was most likely part of the Milky Way disk. In
recent years, however, it has been classified as a high-velocity cloud
because it lies far beyond the main H\,I layer
\citep{Lockman1984,WvW1997}. It has been interpreted variously as a
cloud expelled from the disk \citep{Sofueetal2004},  
or  the gaseous component of the Sgr dwarf spheroidal galaxy 
\citep{Blandhawthornetal1998}.  
We have made an extensive survey of Smith's Cloud in the 21cm H\,I
line using the Green Bank Telescope, whose angular resolution and
sensitivity promised new insights into this system.  In particular, we 
hoped that because of its large angular size we might be able to
measure its transverse velocity (e.g., \citet{Bruns2001,
Lockman2003}). A complete discussion of the observations will appear
elsewhere.  Here we present the initial results on the Cloud's
physical properties and motion.

\section{Observations and Data Reduction}

 Smith's Cloud was observed in the 21cm H\,I line  using the
Robert C. Byrd Green Bank Telescope (GBT) of the NRAO.
  Spectra were measured in both
linear polarizations over a velocity range of  500 km~s$^{-1}$
centered at +50 km~s$^{-1}$ LSR with a channel spacing of 1.03
km~s$^{-1}$ and an effective velocity resolution of 1.25 km~s$^{-1}$.
Spectra were acquired by in-band frequency-switching while moving the telescope
in Galactic latitude or longitude, sampling every $3\arcmin$ in both
coordinates.    In all,  more than 40,000 
positions were measured over an area of $\sim140$ square-degrees. 
Spectra were edited, calibrated, a third-order polynominal was fit to
the emission-free channels, and the data were gridded into a cube with
$3\farcm5$ cell spacing. The rms noise in  the final data cube is
typically 90 mK of brightness temperature in a 1 km~s$^{-1}$ channel.
The on-the-fly mapping and  gridding 
 degraded the  angular resolution somewhat from the intrinsic 
$9\farcm1$ resolution of the telescope to an 
effective resolution of $10\arcmin$ to $11\arcmin$.

\section{Smith's Cloud}

Figure 1 shows the GBT H\,I image of Smith's Cloud at ${\rm V_{LSR} =
100}$ km~s$^{-1}$.  The Cloud has a  cometary morphology
with a bright compact ``tip" and a more diffuse ``tail".  Its
appearance suggests that it is moving toward the Galactic plane at a
$45^{\circ}$ angle and is interacting with the
Galactic ISM.  Direct evidence of this interaction is given in Figure
2, which shows the H\,I emission in velocity-position coordinates along a cut
through the minor axis of the Cloud.   The center of the Cloud has a
velocity near +100 km~s$^{-1}$, well-separated from the Galactic H\,I, but at
its edges the lines are broadened and shifted toward  
the velocity of the Galactic ISM at $\lesssim+35$ km~s$^{-1}$.  We
interpret this as ram-pressure stripping of the Cloud edges. 
The Cloud's interaction with the Galactic ISM is shown further 
in Figure 3, a position-velocity slice 
 along the major axis of the Cloud following a track marked by
the arrows to the upper right and lower left in Fig.~1.  There are 
kinematic bridges between the Cloud and Galactic emission 
(several marked with dashed arrows) as well as 
 clumps of H\,I  (two marked by solid arrows) at
velocities $\lesssim 40$ km~s$^{-1}$ which  correspond to 
gaps in the Cloud.   The clumps are likely material stripped 
from the Cloud. 
 
 \section{Distance to the Cloud}

Portions of Smith's Cloud appear to have been decelerated by the 
ambient medium through which it moves, and we use this to estimate a distance
to the Cloud. The GBT data show disturbances in Galactic H\,I
attributable to the influence of Smith's Cloud at $ V_{\rm LSR} \geq
35$ km~s$^{-1}$ but not at $ V_{\rm LSR} \leq 0$
km~s$^{-1}$.  If Smith's Cloud is interacting with Galactic gas whose
normal rotational velocity is in this range, it implies that that the
Cloud has a distance ${\rm 11.1 < d_{\rm k} < 13.7}$ kpc; 
 the ``far" kinematic distance for a flat rotation curve with
$R_0 = 8.5$ kpc and $V_0 = 220$ km~s$^{-1}$. 

There are other determinations of the distance.  The brightness of
diffuse H$_{\alpha}$ emission from the Cloud and a model
for the Galactic UV flux give either 1 or 13 kpc
\citep{Blandhawthornetal1998, Putmanetal2003}.  Recently 
\citet{Wakkeretal2008} have bracked the distance by looking for the
Cloud in absorption against several stars, finding $ 10.5 < d \leq
14.5$ kpc. The three methods give identical results, 
and we adopt the kinematic distance $d  = 12.4 \pm
1.3$ kpc for the remainder of this Letter.

\section{Properties of the Cloud}

Smith's Cloud lies in the inner Galaxy  below the Perseus spiral arm, 
 $R = 7.6$ kpc from the Galactic center.  
 Properties of the Cloud from the GBT data are
presented in Table 1.  The brightest HI emission at $\ell,b =
38\fdg67$ --13$\fdg41$ is near the Cloud tip.  The H\,I mass of
$10^6$ M$_{\odot}$ is a lower limit because the Cloud appears to
consist of multiple fragments spread over a wide area, not all of
which are covered in the GBT map. It probably contains a significant
mass in H$^{+}$ as well \citep{Wakkeretal2008}.  The H\,I line width $\Delta
v$ (FWHM) varies from $\leq 10$ km~s$^{-1}$ in a band from the Cloud
tip down along the major axis, to $>20$ km~s$^{-1}$ in the Cloud's
tail and at its edge, where the sight-line through the Cloud
intersects regions with a great spread of velocity (e.g., Figs.~2 and
3). The Cloud as a whole is not self-gravitating, and even the
 most compact components have only $1\%$ of their virial mass. 

There are narrow unresolved ridges of intermediate-velocity H\,I 
with N$_{\rm HI} = 2 \times 10^{20}$ cm$^{-2}$ at the
edge of the Cloud (Figure 4, also marked with the arrow in Fig.~2). 
  There is some  cool
gas in the ridge, especially toward the tip of the Cloud, but in general
the lines from the ridge are broad with $\Delta v = 10 - 20$ km~s$^{-1}$.  
The ridge contains orders of
magnitude more gas than would be expected for material swept-up by
the passage of the Cloud through the Galactic halo at a distance of a
few kpc from the plane \citep{DickeyLockman,Howketal2003} but it has
an N$_{\rm HI}$  comparable  to that of the Cloud.  The ridge is most
likely material which has been ram-pressure stripped from the Cloud. 

In the GBT data 
the edge of Smith's Cloud is unresolved along the region of
interaction implying a size $<35$ pc.    The
ridge and the edge of Smith's Cloud do not overlap on the sky, and
both have unresolved edges where they are closest, suggesting that we
are viewing the interaction at a very favorable angle. 
The decellerated clumps marked in Fig.~3 have N$_{\rm HI} = 1-2
\times 10^{19}$ cm$^{-2}$ and H\,I masses of 200 -- 400 M$_{\Sun}$.
The lower-velocity clump is unresolved with a size $<35$ pc, while the
higher-velocity clump is elongated with a size $<35 \times 130$ pc.
The linewidths are narrow, 4 and 5.4 km s$^{-1}$, respectively,
indicating that the clumps contain gas no hotter than 350 K.

\section{Trajectory}

Smith's Cloud covers  a large area on
the sky and we can hope to derive its complete space motion
 from its morphology and the systematic change in
 $V_{\rm LSR}(\ell, b)$ with position if local effects like drag are 
small. For a system centered on the Galactic center, 
\begin{equation}
V_{\rm LSR} = \left[ R_0 sin(\ell) \lbrace{V_{\theta}\over R} - {V_0\over R_0} \rbrace - V_R \cos(\ell+\theta)\right] \cos(b) + V_z sin(b) 
\end{equation} 
where $V_{\theta}$, $V_R$ and $V_z$ are velocity components in the
direction of Galactic rotation, outward, and vertically toward the
north Galactic pole. The angle $\theta$ is measured from the
Sun-center line in the direction of Galactic rotation.  
With Smith's Cloud we have the exceptional circumstance that 
all distances
and angles are known, so with measurements of $V_{\rm{LSR}}(\ell,b)$
and the angle of Cloud motion (assumed to be along its major axis), we
can solve for the individual velocity components and calculate the
total space velocity $V_{tot} \equiv \lbrace {\rm V_{\theta}}^2 +
V_R^2+V_z^2 \rbrace^{\onehalf} $.

The H\,I profiles from Smith's Cloud have complex shapes and a full
discussion of the Cloud kinematics is beyond the scope of this Letter.
As an initial step we have taken the velocity of the brightest
H\,I in each pixel, 
 averaged this over square-degree regions along the major
axis of the Cloud from $(\ell,b) = 36{^\circ}$-$11{^\circ}$ to
$48{^\circ}$-$23{^\circ}$, and solved for the velocity components.  
The results are drawn on Fig.~3.  The main 
Cloud has two kinematic groups, each of
which shows the regular velocity pattern expected from projection
effects of the motion of a coherent object.  Table 2 summarizes the
results, where the quantity $V_{\rm ISM}$ is the velocity of the Cloud
with respect to a corotating ISM at its location, and the
uncertainties reflect both the scatter in the data and in the
$45^{\circ} \pm10^{\circ}$ assumed angle of motion of the Cloud across
the sky.  With ${\rm V_{tot}\approx 300}$ km~s$^{-1}$ the Cloud is
bound to the Galaxy.  Its motion is prograde, somewhat faster than
Galactic rotation, with an outward radial velocity $V_R \sim 100~{\rm
km~s^ {-1}}$. The compact Cloud tip is moving toward the plane with a
velocity  $V_z = 73\pm 26~{\rm km~s^{-1}}$, while the more diffuse trailing 
structure appears to have a much lower vertical velocity of $8 \pm
11~{\rm km~s^{-1}}$.
 
From its current position and velocity we calculate the Cloud's orbit
in the potential of \citet{Wolfire95}.  Assuming that drag is not
significant, it will cross the Galactic plane at a distance $R \approx
11$ kpc from the Galactic center in about 27 Myr.  Retracing its orbit
into the past (again neglecting drag), the Cloud reached
perigalacticon at R$=6.9$ kpc some 12 Myr ago, was never more than 3.6
kpc below the Galactic plane, and actually passed through the plane, 
from above to below, at R=13 kpc about 70 Myr ago. The current orbit 
is tilted only $\approx 30\arcdeg$ to the Galactic pole, so we
probably view the Cloud at a large angle to the plane of the sky, 
with its tail  closer to us than its tip.  These results will likely be 
modified as we understand the Cloud's structure in more detail, but 
the conclusion that large portions of Smith's Cloud move coherently 
seems secure.

\section{Discussion}

All the data on Smith's Cloud are consistent with the model 
 of a $10^6$ M$_{\Sun}$ H\,I Cloud the size of a dwarf galaxy
on  track to intersect the Galactic plane. We know more about this
particular HVC than any other. Its total space velocity of $\approx 300$ 
km~s$^{-1}$ implies that it is bound to the Galaxy and components which
are not greatly slowed by drag from the Galactic halo should reach the
plane in 27 Myr at a location about 11 kpc from the Galactic center.
Its trajectory is rather flat and  mostly prograde, and it may 
have passed through the Galactic plane once before some 70 Myr ago.  
  The Cloud now consists of two coherent kinematic components, as well as
material decellerated to much lower velocity.

Studies of Galactic chemical evolution have uniformly  concluded that
the Milky Way is not a ``closed box" but must accrete 
low-metallicity gas, possibly supplied by HVCs (e.g.,
\citet{Frieletal2002, Matteucci2004, Romanoetal2006, Putman2006}). 
The collision of an HVC with the disk has also been invoked to explain
the largest H\,I supershells as well as Gould's Belt
\citep{TenorioTagle1987,MirabelMorras1990,ComeronTorra1994}.  
While there are HVCs that show evidence of interaction with the Milky
Way halo \citep{BrunsMebold2004} there are only a few known to be 
interacting with the Galactic disk 
\citep{Lockman2003,McClure-Griffithsetal2008} and these are located in the
outer parts of the Galaxy, far from the main star-forming regions.
Smith's Cloud is exceptional in that it is entering the Milky Way at 
$R \lesssim  R_0$.

At its current distance of $\sim 3$ kpc from the Galactic plane the
Cloud is probably encountering a mix of warm H$^+$ and $10^6$ K gas
with a total density in the range $10^{-3}$ to $10^{-4}$ cm$^{-3}$,
though Galactic H\,I clouds at this height are not out of the question
\citep{Lockman2002,Howketal2003,Benjamin2004, Pidopryhoraetal2007}. We
find several Smith Cloud clumps with 
$M_{\rm HI} \approx 100$ M$_{\Sun}$  which
have been decelerated by $>50$ km s$^{-1}$ suggesting that the ambient
ISM is irregular with large density variations. The nature of the
Cloud's interaction with the ISM will depend on its internal
properties which are not yet known, but Smith's Cloud shows every
evidence of being disrupted and may not survive much longer as a
coherent structure.

Does this cloud have a galactic or extragalactic origin? Is Smith's
Cloud a true high-velocity cloud? Its H\,I mass is in the range of
HVCs like Complex C, Complex H, and the Cohen Stream
\citep{Lockman2003, Wakkeretal2007, Wakkeretal2008}, and given its
estimated orbit, it is hard to envision an event that would accelerate
more than $10^{6}~M_{\sun}$ of material to such a high space velocity
with a significant radial component. However, an extragalactic origin
is somewhat problematic as well. If it were extragalactic, it is
puzzling that the orbit is tilted by only $30\arcdeg$ from the
Galactic plane, is prograde, and differs from the ambient ISM by only
130 ${\rm km~s^{-1}}$. Kinematically, then, it might appear to be more
of an intermediate- than a high-velocity cloud, but presumably its
orbit must have been affected by drag well before the current epoch
\citep{BenjaminDanly}.  Its internal kinematics suggest that it has
already fragmented (Fig.~3).  A key observational datum is the Cloud's
metal abudance to establish exactly what kind of gas it is depositing
in the Galaxy.  For this Letter we have analyzed only a fraction of
the information in the GBT H\,I spectra. A more detailed analysis of
the GBT data and additional measurements of this extraordinary and
beautiful object are underway.

\acknowledgments
We thank W.B. Burton for many insightful discussions, and 
 the National Science Foundation for supporting AJH 
through the Research Experience for Undergraduates program.

{\it Facilities:} \facility{GBT}


\begin{figure}
\includegraphics[angle=0,scale=0.5]{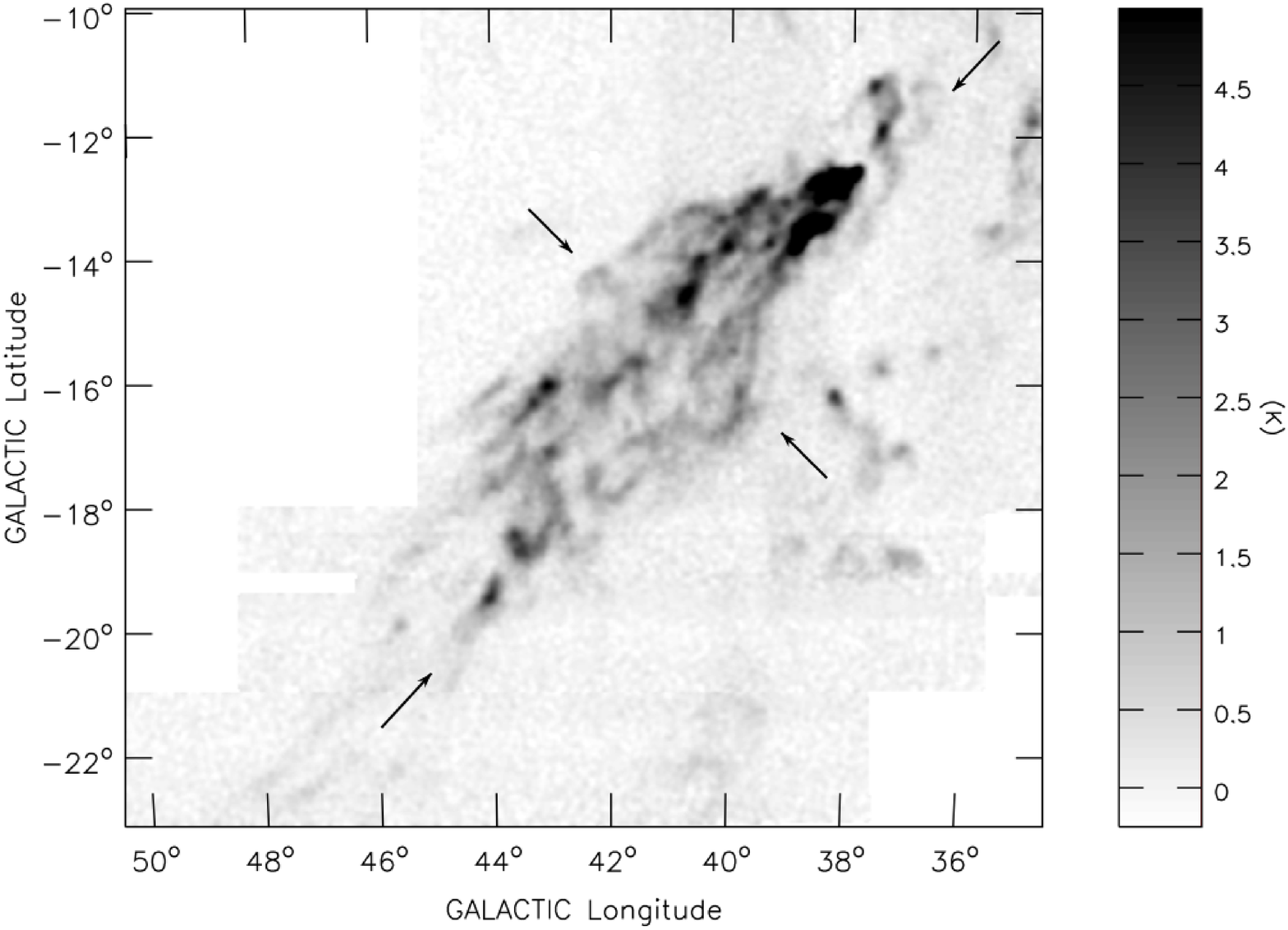}
\caption{GBT H\,I image of Smith's Cloud 
at ${ V_{\rm LSR} = 100}$ km~s$^{-1}$ showing the cometary morphology
strongly suggesting that the Cloud is moving to lower longitude and
towards the plane, and is interacting with the Galactic ISM.  Arrows
mark the tracks of the velocity-position slices of Figs.~2 and 3.}
\end{figure}
\clearpage

\begin{figure}
\includegraphics[angle=0,scale=0.5]{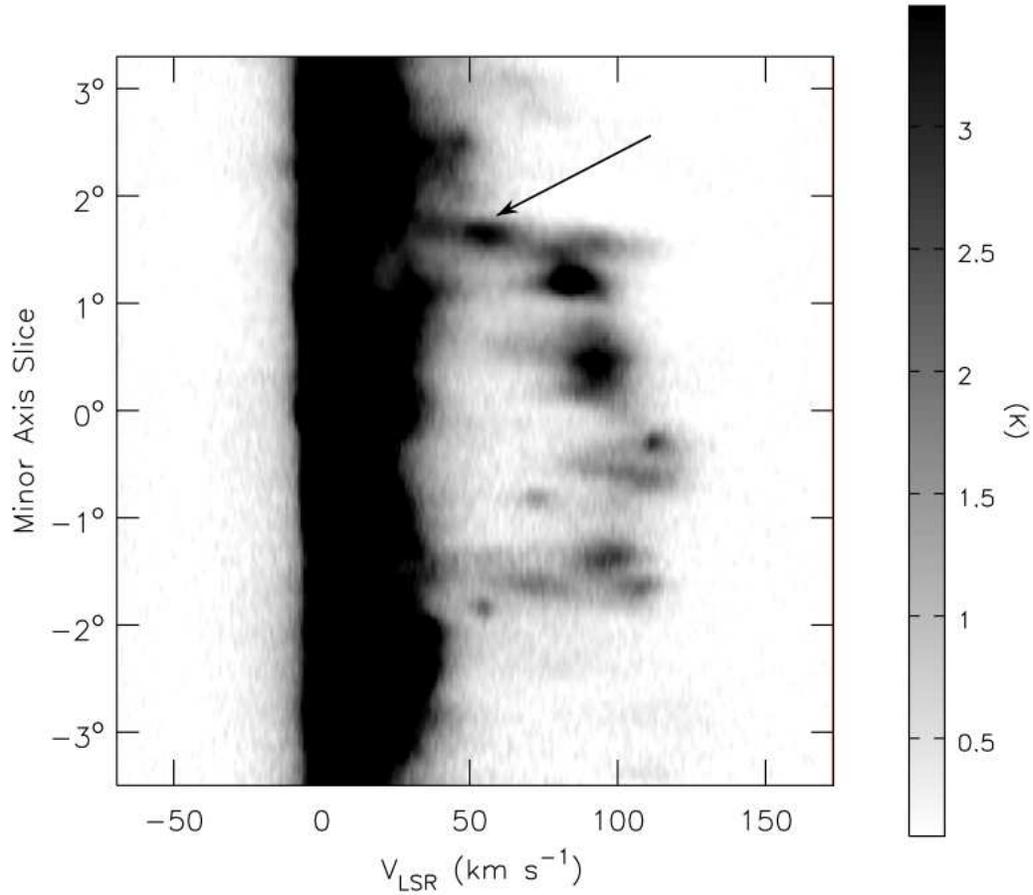}
\caption{GBT H\,I velocity-position slice through Smith's Cloud along a 
track through the minor axis of the Cloud (marked by arrows in 
Fig.~1).  The edges of the Cloud show a sharp gradient in velocity 
 from ${ V_{\rm LSR} \sim100}$ km~s$^{-1}$ 
to the lower velocities of Galactic H\,I.  We interpret this as
evidence of the interaction between the Cloud and the gaseous 
halo of the Milky Way. The arrow marks the decellerated 
ridge shown in Fig.~4.}

\end{figure}
\clearpage

\begin{figure}
\includegraphics[angle=270,scale=0.55]{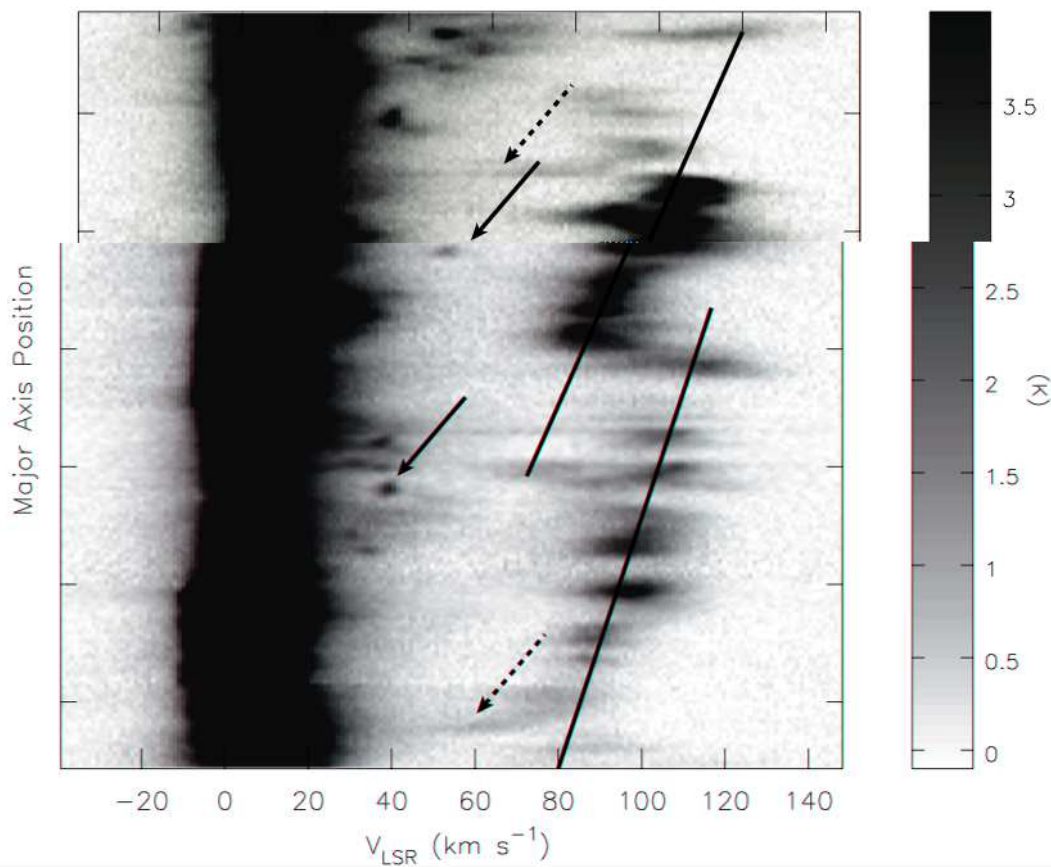}
\caption{GBT H\,I velocity-position slice through 
the major axis of the Cloud at the location of the arrows in Fig.~1.
Marks on the vertical axis are every $157\farcm5$.  Along this track
there are H\,I clumps at low velocity matching gaps in the main
Cloud. The clumps have likely been stripped from the Cloud. Two are
marked by solid arrows.  Two line wings which form kinematic bridges
between the Cloud and Galactic gas are marked with dashed arrows. The
main part of the Cloud shows systematic velocity gradients from the
changing projection of its space velocity with respect to the LSR. The
tilted lines show the expected run of $V_{\rm LSR}$ with position for
${V_{tot}} = 296$ km~s$^{-1}$ (upper part of the Cloud) and ${
V_{tot}} = 271$ km~s$^{-1}$ (lower part).  The Cloud 
consists of at least two coherent kinematic pieces.  }
\end{figure}
\clearpage

\begin{figure}
\includegraphics[angle=0,scale=0.5]{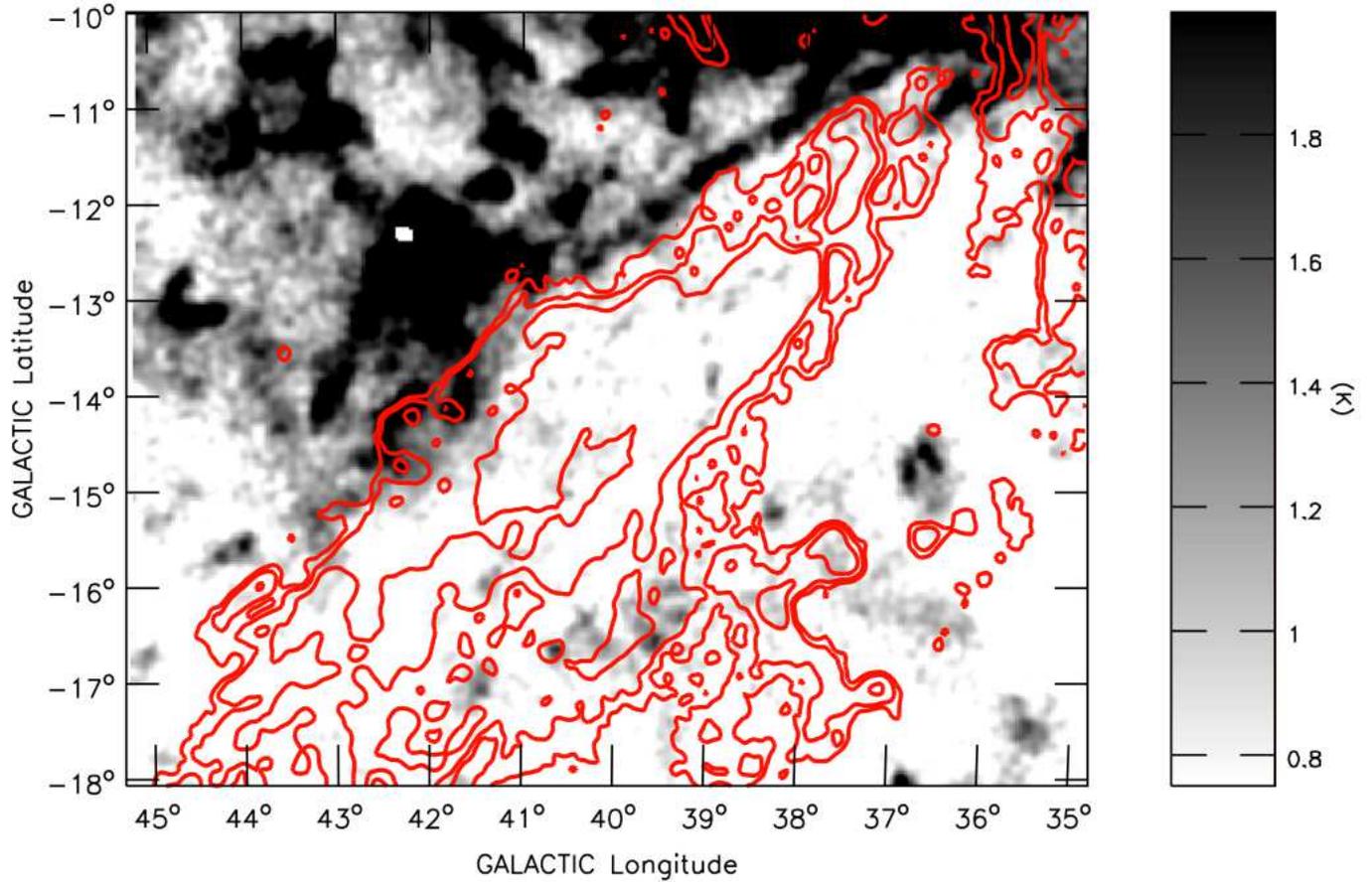}
\caption{H\,I in a 1 km~s$^{-1}$ wide channel at
 51 km s$^{-1}$ (gray scale) with superimposed contours of 
Smith's Cloud showing the ridge of gas resulting from the encounter of
the Cloud with the Galactic halo.  We believe that the lower velocity 
ridge is material stripped from the Cloud.  
The Cloud emission is integrated
over 90 to 145 km s$^{-1}$ and 
contours are drawn at 7, 14 and 35 K km s$^{-1}$.  }

\end{figure}

\begin{deluxetable}{lc}
\tablecolumns{2}
\tablenum{1}
\tablewidth{0pt}
\tablecaption{H\,I Properties of Smith's Cloud}
\tablehead{
\colhead{Property} &
\colhead{Value} \\
\colhead{(1)} & \colhead{(2)} \\
}
\startdata
$\ell, b$  & $38\fdg67 - 13\fdg41 $ \cr
Distance (kpc) & $12.4 \pm 1.3$  \cr
R (kpc) & $7.6\pm0.9$  \\
 z (kpc) & $-2.9\pm0.3$ \\
T$_b$ (K) & 15.5  \\
$\Delta$v (km~s$^{-1}$) & 16.0  \\
 N$_{\rm HI}$ (cm$^{-2}$) & $5.2 \times 10^{20}$  \\
 V$_{\rm LSR}$ (km~s$^{-1}$) & $ 99 \pm 1$  \\
H\,I Mass (M$_{\Sun}$) & $>10^6$  \\
Projected Size  (kpc) & $3 \times 1 $  \cr
\enddata
\tablecomments{All but integral quantities apply to  the direction of greatest
${\rm N_{HI}}$ at the position ${\ell,b} = 38\fdg67$-$13\fdg41$.}
\end{deluxetable}

\begin{deluxetable}{lccccc}
\tablecolumns{6}
\tablenum{2}
\tablewidth{0pt}
\tablecaption{Kinematics of Smith's Cloud}
\tablehead{
\colhead{Location} & \colhead{V$_{\rm R}$} &  \colhead{V$_{\theta}$} & 
\colhead{V$_{\rm z}$} & \colhead{V$_{\rm tot}$} & \colhead{V$_{\rm ISM}$}  \\
\colhead{} & \colhead{(km~s$^{-1}$)}  & \colhead{(km~s$^{-1}$)}  & \colhead{(km~s$^{-1}$)}  & \colhead{(km~s$^{-1}$)}   & \colhead{(km~s$^{-1}$)} \\
\colhead{(1)} & \colhead{(2)}  & \colhead{(3)}  & \colhead{(4)} & \colhead{(5)} 
 & \colhead{(6)} \\ 
}
\startdata
Tip & $94\pm18$ & $270\pm21$ & $73\pm26$ & $296\pm20$ & $130\pm14$ \\
Tail & $129\pm6$ & $220\pm11$ & $8\pm11$ & $271\pm6$ & $130\pm5$ \\
\enddata
\tablecomments{V$_{\rm R}$ is the velocity outward from the Galactic 
Center and V$_{\rm ISM}$ is the total velocity of the Cloud with 
respect to the corotating Galactic ISM at its location.}
\end{deluxetable}

\end{document}